**Pressure-induced giant enhancement of magnetocaloric effects in MnNiSi-based systems**

Tapas Samanta[1], Daniel L. Lepkowski[1], Ahmad Us Saleheen[1], Alok Shankar[1], Joseph Prestigiacomo[1], Igor Dubenko[2], Abdiel Quetz[2], Iain Oswald[3], Gregory T. McCandless[3], Julia Y. Chan[3], Philip W. Adams[1], David P. Young[1], Naushad Ali[2], and Shane Stadler[1]

[1]Department of Physics & Astronomy, Louisiana State University, Baton Rouge, LA 70803 USA
[2]Department of Physics, Southern Illinois University, Carbondale, IL 62901 USA
[3]Department of Chemistry, The University of Texas at Dallas, Richardson, TX 75080 USA

A remarkable decrease of the structural transition temperature of MnNiSi from 1200 K to <300 K by chemically alloying it with MnFeGe results in a coupling of the magnetic and structural transitions, leading to a large magnetocaloric effect near room temperature. It was found that the magnetostructural transition is highly sensitive to external (hydrostatic) pressure: relatively low hydrostatic pressures (~2.4 kbar) lead to an extraordinary enhancement of the isothermal entropy change from $-\Delta S$ = 44 to 89 J/kg K at ambient (atmospheric) and 2.4 kbar applied pressures, respectively, for a field change of $\Delta H$ = 5 T. This giant entropy change is associated with a large relative volume change of about 7% induced by 2.4 kbar applied pressure during the magnetostructural transition. The pressure-enhanced magnetocaloric effects are accompanied by a shift in transition temperature, an effect that may be exploited to tune the transition to the required working temperature, and thereby eliminate the need for a given material to possess a large magnetocaloric effect (i.e., entropy change) over a wide temperature range. Furthermore, this material also possesses negligible hysteresis losses.

Magnetic refrigeration techniques based on magnetocaloric effects (MCE) are considered a preferred alternative to the more common, gas-compression-based refrigeration, and are expected to be employed in future solid-state-based refrigeration devices for near-room-temperature applications [1, 2]. The current challenge is to discover and develop materials that exhibit giant MCEs, and to develop mechanisms that improve the MCEs of the refrigerant materials in the context of applications. Until now, only a few classes of materials, such as $Gd_5Si_2Ge_2$ [3], MnAs-based compounds [4, 5], $La(Fe_{1-x}Si_x)_{13}$ [6], MnCoGe-based compounds [7, 8], $Ni_2MnGa$-based Heusler alloys [9], and $Ni_2MnIn$-based Heusler alloys [10], show giant MCEs close to room temperature. The effects are associated with a strong coupling of magnetic and structural degrees of freedom that result in a giant MCE in the vicinity of the magnetostructural transition (MST), accompanied by changes in crystal symmetry or volume. It is highly desirable to not only discover new giant MCE materials, but to find such materials in which these properties can be tuned or enhanced by some other external parameter in addition to magnetic field, such as chemical alloying, pressure, or electric field.

Pressure is a controllable external parameter that can affect the structural entropy change ($\Delta S_{st}$) of a system, where $\Delta S_{st}$ is related to the total entropy change ($\Delta S_{tot}$) and the magnetic entropy change ($\Delta S_M$) through $\Delta S_{tot} = \Delta S_M + \Delta S_{st}$ [11]. However, MnAs is probably the only previously reported system that exhibits a large pressure-induced enhancement of the MCE associated with a large lattice contribution to $\Delta S_{tot}$ as a result of strong magnetoelastic coupling [5]. A hydrostatic-pressure study (barocaloric effect) on a Ni–Mn–In based Heusler alloy also indicates the possibility of applying pressure for improving the magnetocaloric effect by demonstrating a large isothermal entropy change [12]. Recently, a strain-mediated, giant extrinsic MCE has been reported on $La_{0.7}Ca_{0.3}MnO_3$ films associated with the induced strain from a first-order structural transition that occurs in the $BaTiO_3$ substrate [13]. The phenomenon can be explained as a culmination of the usual field-induced MCE effects at ambient pressure associated with the MST together with additional pressure-induced modifications of the structural entropy change. The latter can be viewed as the solid state analogy to conventional cooling technology based on compression (pressurized solid state system) and expansion (unpressurized solid state system) of gases. In this Letter, we demonstrate an extraordinary pressure-induced two-fold enhancement of the total isothermal entropy change from 44 J/kg K at ambient pressure to 89 J/kg K at 2.4 kbar in a relatively unexplored MnNiSi-based system. The large effect is associated with a large relative volume change about 7% as a result of the MST. The applied pressure also shifts the MST to lower temperature, a property that may of utility in devices.

In recent years, there has been much interest in MnTX-based (T=Co, Ni and X=Ge, Si) intermetallic compounds due to the temperature-induced magnetostructural transitions that are responsible for their shape-memory phenomena, giant magnetocaloric effects, and volume anomalies near room temperature [7, 8, 14, 15]. However, instead of a coupled MST, the parent MnTX compounds exhibit a second order magnetic transition at temperatures below the structural transitions and undergo structural transitions from a low-temperature orthorhombic TiNiSi-type structure to a high-temperature hexagonal $Ni_2In$-type structure in the paramagnetic state [16]. Considerable attention has been given to the MnCoGe-based system regarding magnetocaloric effects due to the closely-spaced magnetic ($T_C \sim 355$ K) and structural transitions ($T_M \sim 372$ K) [16], and the potential to couple them through a single substitution, with the hope that the resulting MST will exhibit a large MCE at room temperature [7, 17].

The MnNiSi system is quite different than the abovementioned compounds, in that it exhibits a structural transition at an extremely high temperature of about 1200 K, and $T_C = 662$ K. Therefore, reducing the structural transition at $T_M$ so drastically in order to locate the MST near room temperature is a challenging task for which a single-element substitution is not sufficient. In the present study, it was found that isostructurally alloying MnNiSi with MnFeGe (which has a stable hexagonal structure with $T_C \sim 159$ K) stabilizes the hexagonal $Ni_2In$-type phase by sharply reducing the structural transition temperature from 1200 K to less than 300 K. As a result, coupled magnetostructural transitions have been realized in $(MnNiSi)_{1-x}(MnFeGe)_x$ near room temperature.

Polycrystalline $(MnNiSi)_{1-x}(MnFeGe)_x$ (x=0.52 and 0.54) samples were prepared by arc-melting the constituent elements of purity better than 99.99% in an ultra-high purity argon atmosphere. The samples were annealed under high vacuum for 3 days at 750°C followed by quenching in cold water. The crystal structures of the samples were determined using a room temperature X-ray diffractometer (XRD) employing Cu $K\alpha$ radiation. Temperature-dependent XRD measurements were conducted on a Bruker D8 Advance diffractometer using a Cu $K\alpha_1$ radiation source ($\lambda = 1.54060$ Å) equipped with a LYNXEYE XE detector. A superconducting quantum interference device magnetometer (SQUID, Quantum Design MPMS) was used to measure the magnetization of $(MnNiSi)_{1-x}(MnFeGe)_x$ within the temperature interval of 10-350 K, and in applied magnetic fields up to 5 T. Magnetic measurements under hydrostatic pressure were performed in a commercial BeCu cylindrical pressure cell (Quantum Design). Daphne 7373 oil was used as the pressure transmitting medium. The value of the applied pressure was calibrated by measuring the shift of the superconducting transition temperature of Sn used as a reference manometer (critical temperature $T_C \sim 3.72$ K at ambient pressure) [18]. Heat capacity measurements were performed using a physical properties measurement system (PPMS by Quantum Design, INC) in a temperature range of 220 - 270 and in fields up to 5 T.

The room temperature X-ray diffraction (XRD) patterns of $(MnNiSi)_{1-x}(MnFeGe)_x$ with x=0.52 and 0.54 are shown in Fig. 1. A hexagonal $Ni_2In$-type crystal structure has been detected at room temperature for both compositions. Structural refinement of the XRD data was carried out using the Rietveld profile refinement method, and the lattice parameters at room temperature were found to be $a = 4.082$ Å (4.092 Å), and $c = 5.294$ Å (5.305 Å), for $x = 0.52$ (0.54), respectively. The larger lattice parameters for $x = 0.54$ compared to that of $x = 0.52$ apparently indicates a stabilization of the hexagonal phase at a lower temperature for $x = 0.52$ due to the smaller Mn-Mn separation in the isostructural MnCoGe-based system in its hexagonal phase [19, 20]. However, a closer inspection of the $c/a$ ratio suggests that $c/a$ for $x = 0.54$ is slightly smaller than for $x = 0.52$. In the orthorhombic crystal structure, the reduction of the lattice parameter $a_{ortho}$ can distort the geometry of the crystal structure in orthorhombic MnNiSi and, as a result, the hexagonal crystal structure is stabilized [20]. Since the abovementioned orthorhombic lattice parameter is related to the hexagonal lattice parameter ($a_{ortho} = c_{hex}$), it is expected that a decrease in the $c/a$ ratio tends to stabilize the hexagonal structure. As a result, a decrease in the structural transition temperature has been observed with increasing $x$, and is associated with a reduction of the $c/a$ ratio (see Fig. 2). A similar trend has been reported previously in the isostructural MnCoGe-based system [21]. Furthermore, the temperature-dependent XRD measurements have been performed to estimate the relative volume change associated with the structural transition for $x = 0.54$ (see Fig. 1). A large volume change about 3.3% in the unit cell volume occurs due to the structural transition from the high-temperature hexagonal phase to the low-temperature orthorhombic phase.

The temperature-dependent magnetization ($M$) data at ambient pressure for $(MnNiSi)_{1-x}(MnFeGe)_x$ ($x = 0.52$, 0.54), as well as under the application of hydrostatic pressure for $x = 0.54$, measured during heating and cooling in the presence of a 1 kOe magnetic field, are shown in Fig. 2. A sharp change in magnetization was observed in the vicinity of the phase transition, representing a magnetic transition from a low-temperature ferromagnetic (FM) state to a high-temperature paramagnetic (PM) state. The observed thermal hysteresis between heating and cooling curves indicates that the magnetic and structural transitions coincide, leading to a single first-order MST (at $T_M$) from a FM to a PM state. Increasing the level of substitution of hexagonal MnFeGe shifts $T_M$ to lower temperature while maintaining the coupled nature of the MST. It should be noted that this coupling is maitained only in a very narrow range of concentrations ($0.50 < x < 0.56$).

The application of hydrostatic pressure ($P$) also stabilizes the hexagonal phase at lower temperature, at a rate of decrease $dT_M/dP = $ -4.5 K/kbar for the sample with $x = 0.54$ (see Fig. 2). This shift is possibly associated with a distortion of the orthorhombic lattice that increases the stability of the hexagonal phase. The low temperature $M(H)$ curves as measured at 10 K show a shape typical for FM-type ordering (Fig. 3). The value of the magnetization for 5 T ($M_{5T}$) slightly decreases with increasing $x$. However, the pressure-induced change in $M_{5T}$ is almost negligible, suggesting a minor variation of the FM exchange in the low-temperature orthorhombic phase that may be attributed to a slight modification of the electronic density of states at the Fermi level.

The maximum field-induced entropy change (-$\Delta S$) has been estimated using both the Maxwell relation as well as the Clausius-Clapeyron equation. The temperature dependence of -$\Delta S$ as estimated using the Maxwell relation for the magnetic field change $\Delta H$=1-5 T are plotted in Fig. 4 for the compositions with $x = 0.52$ and 0.54, and was calculated using the isothermal magnetization curves measured at different constant temperatures. A large value of -$\Delta S$ has been detected at ambient pressure associated with the first-order MST. Considering the higher degree of applicability (and reliability) of the Clausius-Clapeyron equation in the vicinity of discontinuous, first-order MST's, the maximum value of $-\Delta S$ also has been estimated using Clausius-Clapeyron equation, yielding a value of 42 J/kg K for $\Delta H$ = 5 T (see Fig. 5). The values of $-\Delta S$ are in good agreement as estimated using the two different equations, which lends justification to the use of the Maxwell relation in the case of this system. The most remarkable observation is that the application of relatively low hydrostatic pressure (~2.4 kbar) leads to a giant enhancement of -$\Delta S$, from ~44 J/kg K (ambient pressure) to 89 J/kg K (P=2.4 kbar), for a field change of 5 T (for x=0.54). It is also important to note the shift of the $T_M$ to lower temperature by 4.5 K/kbar with applied pressure, since it suggests a destabilization of the low-temperature phase, and also reveals a method in which the transition can be tuned in temperature. Moreover, the field-dependent hysteresis loss is negligible in this system.

To estimate the value of -$\Delta S$ as well as the adiabatic temperature change ($\Delta T_{ad}$) at ambient pressure, temperature-dependent heat capacity measurements at various constant magnetic fields have been performed and are shown in Fig. 6. However, the estimation of -$\Delta S$ and $\Delta T_{ad}$ are quantitatively unreliable due to a decoupling of the sample from the heat capacity measurement platform. The bulk polycrystalline sample suffers a structural breakdown into a powder as a result of the drastic structural changes at the MST. The coupling issues notwithstanding, the heat capacity measurements are in qualitative agreement with the magnetization data in terms of the phase transition, but likely underestimate the value of -$\Delta S$ and $\Delta T_{ad}$.

This observed degree of enhancement of -$\Delta S$ is rare and, until now, has only been observed in a MnAs-based system as a result of magnetoelastic coupling [5]. The maximum magnitude of -$\Delta S$ reaches a value of 89 J/kg K with the application of 2.4 kbar for $\Delta H$=5 T, which greatly exceeds that observed in other well-known giant magnetocaloric materials. In this case, the combined effect of pressure and magnetic field could facilitate an improvement in the magnetocaloric working efficiency of the material. As the hydrostatic pressure increases, $T_M$ decreases, and the maximum value of -$\Delta S$ increases in a nearly linear fashion up to 2.4 kbar. A careful examination of the pressure-induced -$\Delta S(T)$ curves indicates that the shape of the -$\Delta S(T)$ curve changes with increasing pressure.

Interestingly, the total area under the $\Delta S(T)$ curve remains nearly constant with application of pressure, as shown in Fig. 7(a). This type of area conservation is in accordance with the maximum limit of the refrigerating power, $\int_0^\infty \Delta S \mathrm{d}T = -M_S * \Delta H$ (where $M_S$ is the saturation magnetization), which is expected to be constant provided $M_S$ remains unchanged ($M \sim 110$ emu/g at $T = 10$ K for $H = 5$ T at ambient pressure, as well as under the condition of applied pressure for $x = 0.54$, see Fig. 3). Therefore, the decrease in the width of the -$\Delta S(T)$ curve is compensated by an increase in its maximum value as the pressure increases.

For a MST, the total entropy change ($\Delta S_{tot}$) can be expressed as $\Delta S_{tot} = \Delta S_M + \Delta S_{st}$, where $\Delta S_M$ and $\Delta S_{st}$ are the magnetic and structural entropy changes, respectively [11]. To understand the origin of the observed giant enhancement of the MCE, the relative influence of both $\Delta S_M$ and $\Delta S_{st}$ on the total entropy change ($\Delta S_{tot}$) associated with the MST has been investigated. Previous MCE studies based on the isostructural MnCoGe system indicate that the structural entropy change is much more significant than the magnetic entropy change in this class of materials [11]. It is expected that the application of external hydrostatic pressure predominantly affects the structural transition. Moreover, the FM Curie temperatures of both the martensitic orthorhombic and the austenite hexagonal phases remain largely unchanged, as is evident from earlier studies on the isostructural MnCoGe system, where $T_M$ shows a strong composition dependence [16]. This suggests a possible weak variation of the FM exchange in both phases, indicating a minor alteration of the electronic density of states at the Fermi level. As a result, we can assume that the application of hydrostatic pressure may have a similar effect on the density of states in our system.

For the first-order magnetic phase transitions, the Clausius-Clapeyron thermodynamic relation states that $(\mathrm{d}T_M/\mathrm{d}P) = -(\Delta V/\Delta M)(\mathrm{d}T_M/\mathrm{d}H)$, where $\Delta V$ is the change in volume due to the structural transition from the martensitic orthorhombic to the austenitic hexagonal phase in the vicinity of the MST. Considering that the maximum entropy change according to the Clausius-Clapeyron equation is $\Delta S = (\Delta M/\Delta T)\Delta H$, the previous expression can be written as $(dT_M/dP) = -(\Delta V/\Delta S)$. As the shift in $T_M$ varies almost linearly with pressure $(dT_M/dP = -4.5$ K/kbar$)$, a proportional relationship between $\Delta S$ and $\Delta V$, i.e., $\Delta S \propto \Delta V$, can be expected in this case. As mentioned before, the values of $-\Delta S$ as estimated using either the Maxwell relation or the Clausius-Clapeyron equation are in good agreement. Therefore, the observed pressure-induced, two-fold increase of $|-\Delta S|$ from 44 to 89 J/kg K is associated with a large volume change during the MST from a FM orthorhombic to a PM hexagonal phase. We can now apply the relationship between $\Delta S_{st}$ and the relative volume change as $\delta(\Delta V/V(\%))/\delta(\Delta S_{st})=0.08$ (J/kg K)$^{-1}$. The relative change in $\Delta V$ that results in this type of giant enhancement of $\Delta S$ can be estimated by keeping in mind that $\Delta S_M$ is usually very small compared to $\Delta S_{st}$, as previously observed in this type of system [11]. A graphical comparison has been made (shown in Fig. 7(b)), which indicates that the application of 2.4 kbar of pressure induces a relative volume change of $\Delta V/V \sim 7.1\%$, and results in an enormous increase in $\Delta S$. Based on the result of the structurally-driven giant MCE reported in Ref.[10], it could be expected that a pressure-induced giant enhancement of $\Delta S$, as a result of a large increase of $\Delta S_{st}$ would effectively increase the value of $\Delta T_{ad}$ in the presently studied system. It is worth pointing out here that the relative volume change at ambient pressure as estimated from the graphical comparison in Fig. 7(b) ($\sim 3.5\%$) is in good agreement with the result from the temperature-dependent XRD measurements ($\Delta V/V \sim 3.3\%$).

As demonstrated in this study, hydrostatic pressure acts as a parameter that leads to a giant enhancement of the magnetocaloric effect in $(MnNiSi)_{1-x}(MnFeGe)_x$, and is associated with an extreme volume change ($\sim 7\%$) in the vicinity of the MST. The pressure-induced volume change during the MST significantly enhances the structural entropy change, and results in a giant enhancement of the total isothermal entropy change by about two-fold, from 44 J/kg K at ambient pressure to 89 J/kg K at $P = 2.4$ kbar. The applied pressure also shifts the transition to lower temperature, a behavior that may be exploited in devices. Finally, this largely unexplored system has additional advantage such as negligible hysteresis losses, and nontoxic constituent elements, which makes this MnNiSi-based system a promising candidate for room-temperature magnetic refrigeration applications.

**Acknowledgements**

Work at Louisiana State University (S. Stadler) was supported by the U.S. Department of Energy (DOE), Office of Science, Basic Energy Sciences (BES) under Award No. DE-FG02-13ER46846, and heat capacity measurements were carried out at LSU by P. W. Adams who is supported by DOE, Office of Science, BES under Award No.  DE-FG02-07ER46420. Work at Southern Illinois University was supported by DOE, Office of Science, BES under Award No. DE-FG02-13ER46946. D. P. Young fabricated samples and acknowledges support from the NSF through DMR Grant No. 1306392. XRD measurements were carried out by J. Y. Chan who was supported by NSF under DMR Grant No.1358975.

**Figure Captions:**

**Fig. 1** (Color online). Temperature-dependent XRD patterns for $(MnNiSi)_{1-x}(MnFeGe)_x$ with $x = 0.54$ and room temperature XRD for $x = 0.52$. The Miller indices of the hexagonal and orthorhombic phases are designated with and without an asterisk (*), respectively.

**Fig. 2** (Color online). Temperature dependence of the magnetization in the presence of a 1 kOe magnetic field during heating and cooling for $(MnNiSi)_{1-x}(MnFeGe)_x$ as measured at ambient pressure for both compositions, and at different applied hydrostatic pressures for $x = 0.54$. Arrows indicate the direction of change of temperature along the $M(T)$ curves.

**Fig. 3** (Color online). Isothermal magnetization curves as a function of magnetic field ($H$) at $T = 10$ K for $(MnNiSi)_{1-x}(MnFeGe)_x$.

**Fig. 4** (Color online). Plot of the total isothermal entropy changes ($-\Delta S$) as a function of temperature for different magnetic field changes of $\Delta H$=1-5 T. The values were calculated using the Maxwell relation at ambient pressure for both compositions ($x = 0.52$ and $0.54$), as well as under conditions of different applied hydrostatic pressures for $x = 0.54$.

**Fig. 5** (Color online). The heating thermomagnetization curves for applied fields $H$ =0.1 and 5 T used to estimate the value of $-\Delta S$ for $x = 0.54$ with the Clausius-Clapeyron equation.

**Fig. 6** (Color online). Heat capacity ($C_P$) as a function of temperature for $x = 0.54$ at different constant magnetic fields.

**Fig. 7** (Color online). (a) Composition-dependent total integrals showing the area under the $\Delta S(T)$ curves at ambient pressure, and as a function of applied hydrostatic pressure (for $x = 0.54$). (b) Linear dependence of the relative volume changes ($\Delta V/V$) (left axis) with the structural entropy changes ($\Delta S_{st}$) from [Ref. 11] plotted as a dashed line. Pressure-induced modification of $\Delta S_{tot}$ for $x = 0.54$ (solid symbols) considering $\Delta S_{tot} \sim \Delta S_{st}$ since $\Delta S_{st} >> \Delta S_{M}$.

Fig1

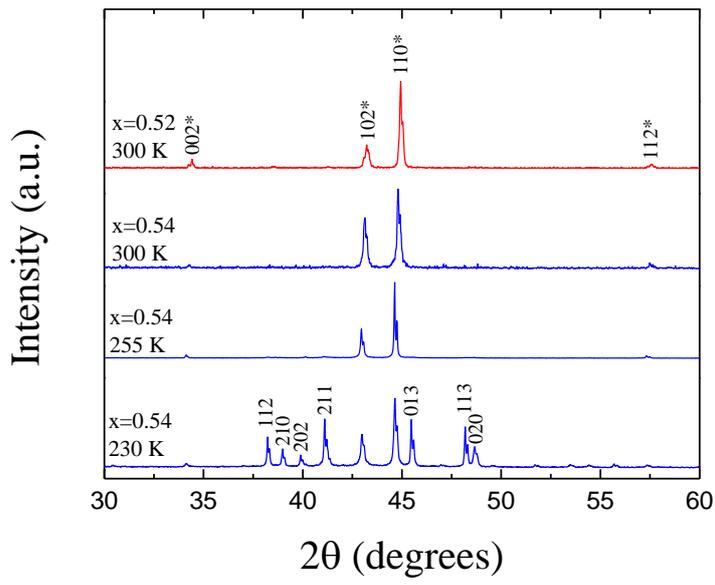

Fig2

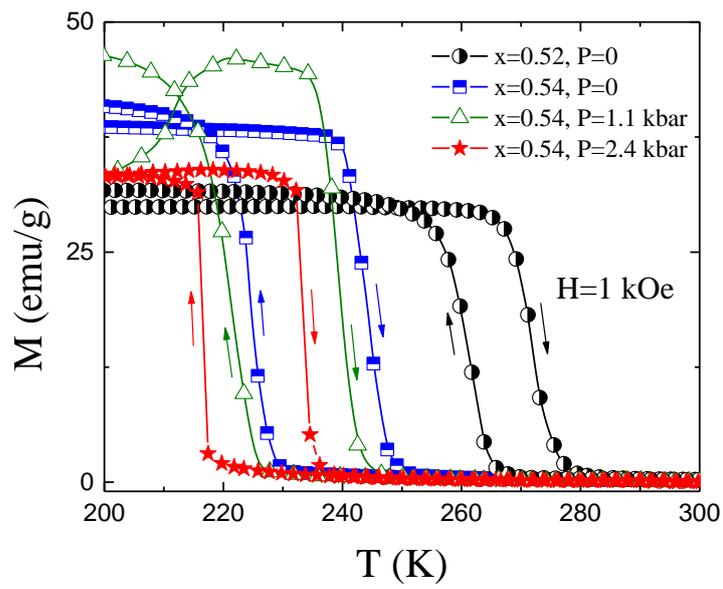

Fig3

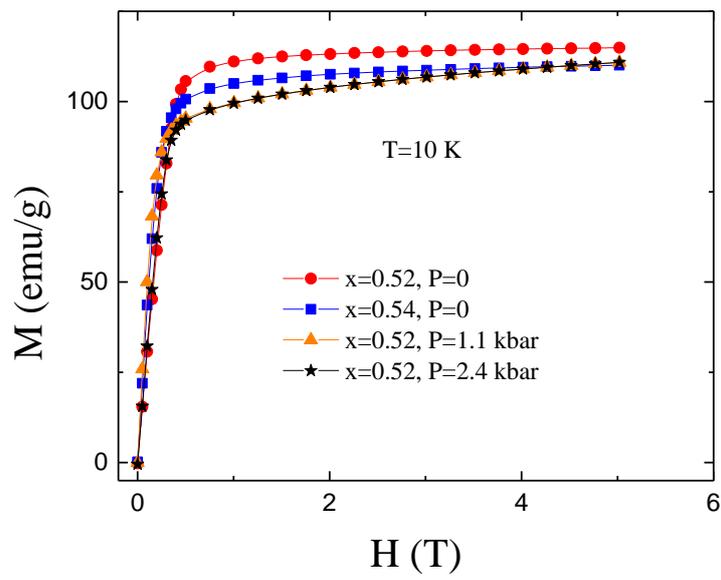

Fig4

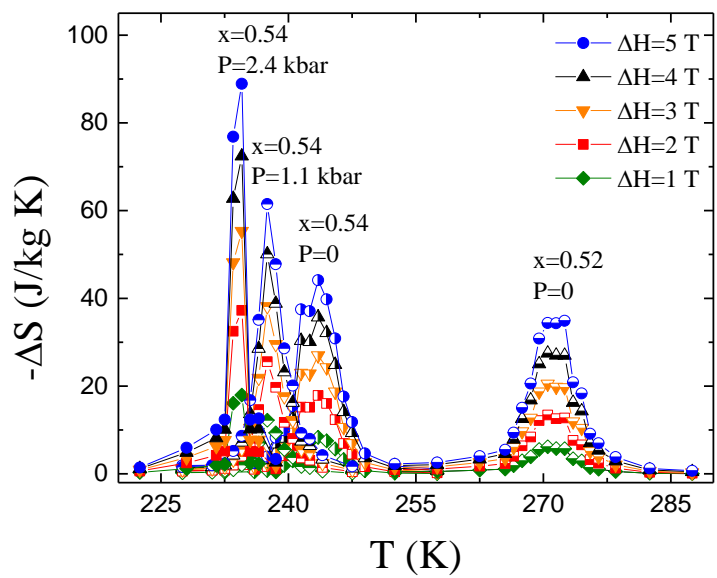

Fig5

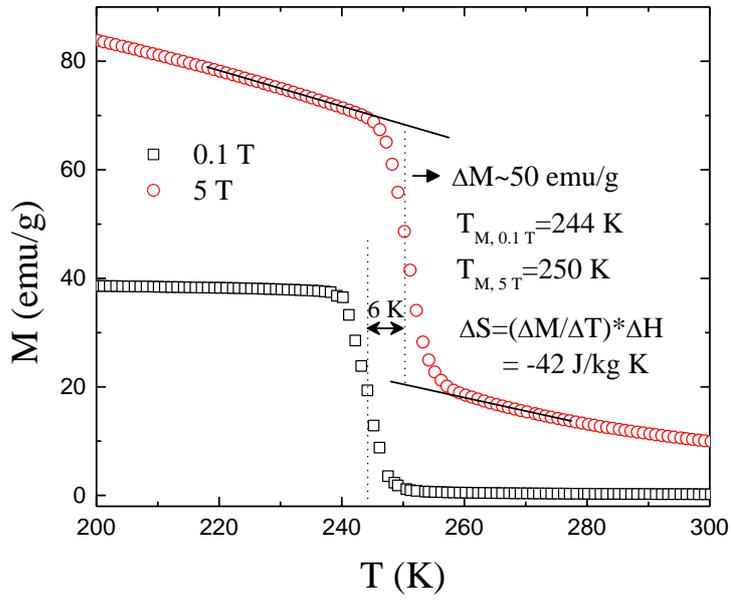

Fig6

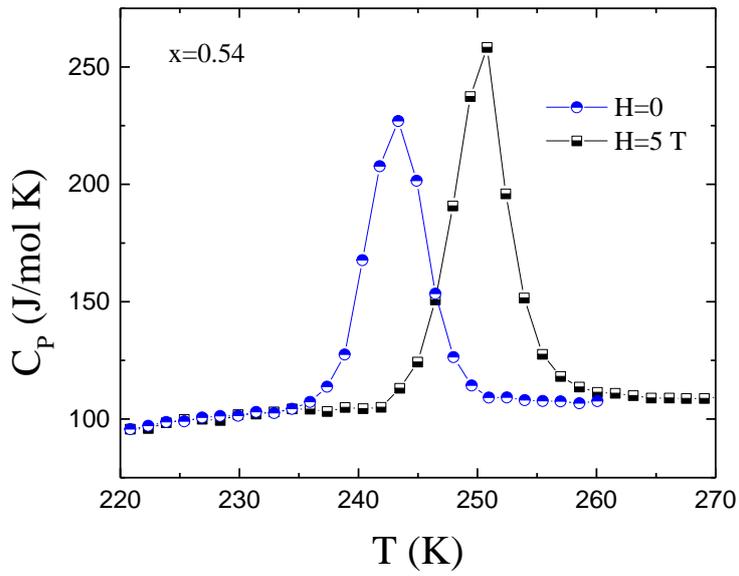

Fig7

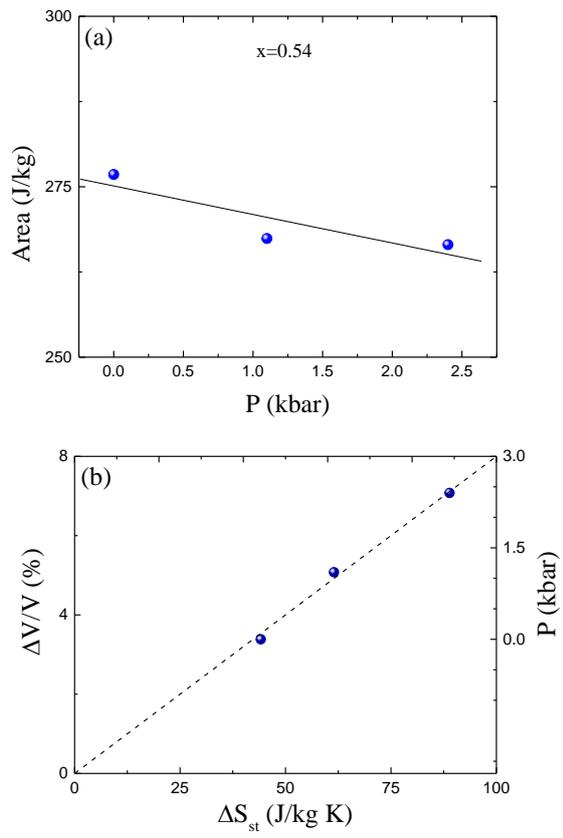

(a)  x=0.54

Area (J/kg)

P (kbar)

(b)

ΔV/V (%)

P (kbar)

ΔS$_{st}$ (J/kg K)